# Linking Order to Strength in Metals

Nicolas Argibay[1,*], Duane D. Johnson[1,2], Michael Chandross[3], Ryan T. Ott[1], Hailong Huang[1], Rameshwari Naorem[1], Gaoyuan Ouyang[1], Andrey V. Smirnov[1] and Prashant Singh[1]

[1] *Division of Materials Science and Engineering, Ames National Laboratory, IA*
[2] *Department of Materials Science and Engineering, Iowa State University, IA*
[3] *Material, Physical, and Chemical Sciences Center, Sandia National Laboratories, NM*
[*]Corresponding author: nargibay@ameslab.gov

Keywords: strength; structural ordering; chemical ordering; high-entropy alloys (HEAs); multi-principal-element alloys (MPEAs); complex-concentrated alloys (CCAs); intermetallic; metallic glass; amorphous

*Abstract*

The metallurgy and materials communities have long known and exploited fundamental links between chemical and structural ordering in metallic solids and their mechanical properties. The highest reported strength achievable through the combination of multiple metals (alloying) has rapidly climbed and given rise to new classifications of materials with extraordinary properties. Metallic glasses and high-entropy alloys are two limiting examples of how tailored order can be used to manipulate mechanical behavior. Here, we show that the complex electronic-structure mechanisms governing the peak strength of alloys and pure metals can be reduced to a few physically-meaningful parameters based on their atomic arrangements and used (with no fitting parameters) to predict the maximum strength of any metallic solid, regardless of degree of structural or chemical ordering. Predictions of maximum strength based on the activation energy for a stress-driven phase transition to an amorphous state is shown to accurately describe the breakdown in Hall-Petch behavior at the smallest crystallite sizes for pure metals, intermetallic compounds, metallic glasses, and high-entropy alloys. This activation energy is also shown to be directly proportional to interstitial (electronic) charge density, which is a good predictor of ductility, stiffness (moduli), and phase stability in high-entropy alloys, and in solid metals generally. The proposed framework suggests the possibility of coupling ordering and intrinsic strength to mechanisms like dislocation nucleation, hydrogen embrittlement, and transport properties. It additionally opens the prospect for greatly accelerated structural materials design and development to address materials challenges limiting more sustainable and efficient use of energy.

*Introduction*

As context for a generalized understanding of how order is linked to maximum or intrinsic strength, it is helpful to begin with a discussion of notable examples of where limits in the experimental strength of metals were reported. In the late 1980s and early 1990s, Gleiter and others showed that the increase in strength with decreasing grain size in metals breaks down at a fine enough grain size (~10 nm), below which a plateau or softening is typically observed [1,2]. This breakdown in Hall-Petch behavior was first given a physical basis by Conrad and Narayan [3], who attributed it to a crossover in energetic favorability from dislocation emission to stress-assisted diffusive transport along grain boundaries. However, this model is not predictive as it requires a-priori knowledge of the activation volume, the origin of which was unclear. A key premise in their interpretation is also problematic in some cases, in that it implies that the rate of deformation under grain-boundary sliding is diffusion limited. Molecular-dynamics (MD) simulations [4–6] and experiments [7,8] have since shown that this behavior persists at extremely high-deformation rates and low temperatures, regimes where deformation rates far exceed those that can be accommodated via solid diffusion.

Recently, a predictive model with no fitting parameters was developed to explain the strength regime associated with grain sizes below a critical value where Hall-Petch breakdown occurs. This model is based on the free energy change per atom to transform coordination or bond density from crystalline to liquid-like [9]. The model was later extended to successfully describe the temperature-dependent strength of bulk-metallic glasses in the inhomogeneous flow regime, i.e., below their glass-transition temperature at strain rates that obviate solid diffusion-limited flow [10]. The activation energy for flow (i.e., flow strength) was shown to be a function of changes in average coordination number and the size of atomic clusters (short-range order), avoiding the fitting parameter in the work by Spaepen based on a similar approach [11]. More recent work showed that it is possible to exploit low-free-energy microstructural configurations (e.g., nanoscale coherent twins) to suppress dislocation-mediated plasticity mechanisms and attain the predicted strength maxima in bulk nanostructured metals [12,13]. These results confirmed the physical basis for the early formulation of the strength framework for pure metals and dilute alloys at the nanoscale limit [9], where a simplifying assumption of high-energy (i.e., high structural disorder) grain boundaries was made. Recent work has shown that the stiffness and limiting or ultimate strength of metals, including concentrated solid-solution and high-entropy alloys, can be reduced to a single physical parameter, i.e., the average interstitial charge density [14]. This discovery, which offers a greatly reduced computational cost for alloy design, showed that the stiffness (moduli) and ultimate strengths of chemically-disordered, structurally-ordered alloys are proportional and can be predicted by a rule-of-mixtures of tabulated elemental values of interstitial charge density for their lattice structure in the alloyed state.

Here we present a general framework for predicting the maximum strength of metals based the stress-activated formation of an amorphous interface in initially variably-ordered (chemically and structurally) metallic solids. We also show that computationally-inexpensive methods that determine material properties like enthalpy and entropy of mixing can be combined with elemental properties and structural information, like grain size and related structural order parameters, to accurately predict the maximum strength of a broad range of alloy systems. These predictions are shown to be in good agreement with those based on interstitial-charge density for random solid solutions. This link establishes a pathway for assessing phase stability [15], as well as ductility by comparing activation energies for competing mechanisms of slip (e.g., unstable stacking-fault energy) and fracture (i.e., free-surface formation energy).

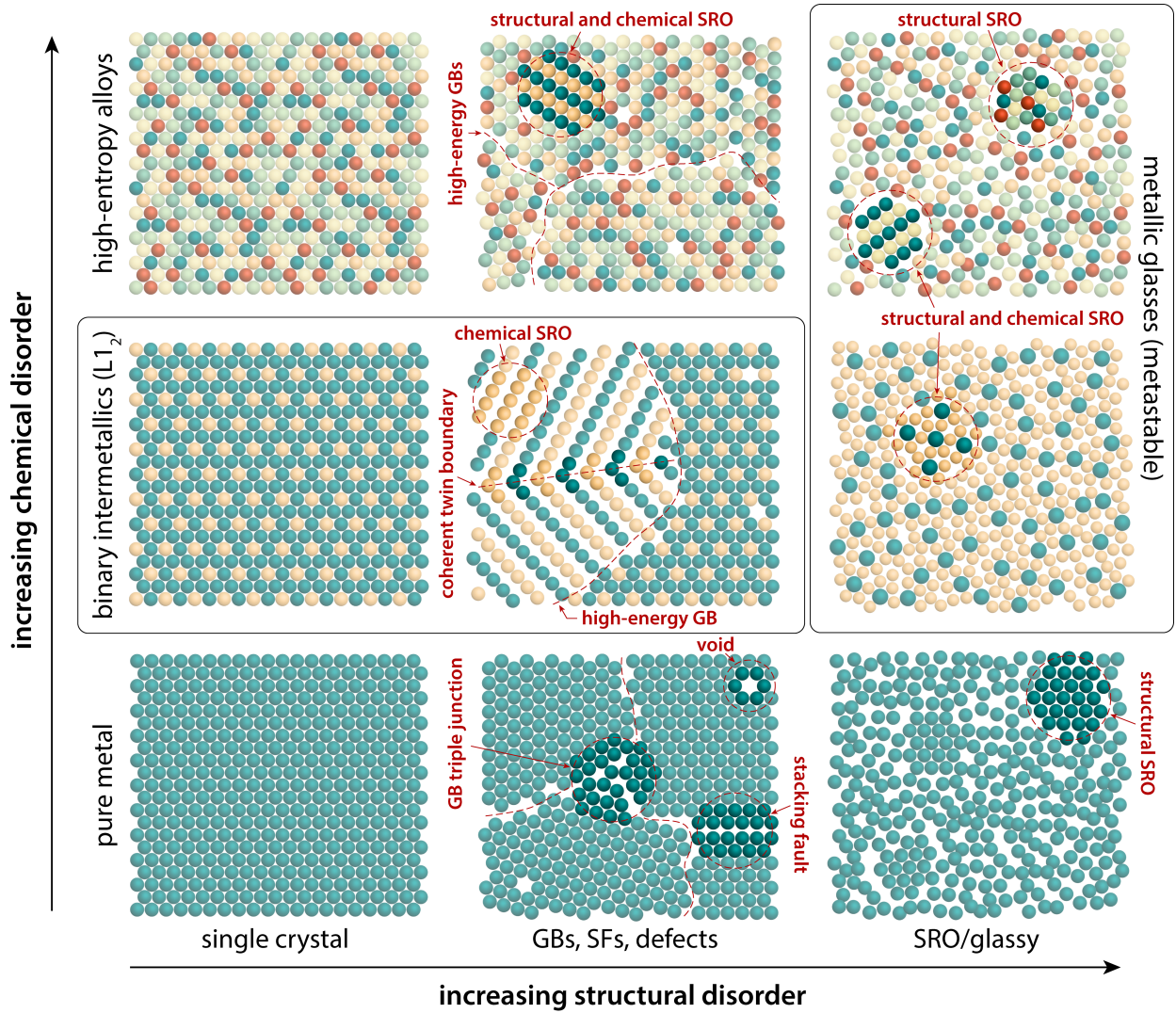

*Figure 1.* Illustration of structural and chemical disorder in metals.

## I. A Free-Energy-Based Framework for Order-Dependent Strength of Metals

We consider the deformation of pure metals to be an *activated process*, based on changes in the free energy of the system that depend on temperature and strain rate. This approach, relying exclusively on materials properties with no fitting parameters, was first developed for polycrystalline pure metals [9] and later for metallic glasses [10]. The fundamental basis stems from earlier work by Arrhenius, Eyring [16], and Prandtl [17] on chemical-reaction rates and shear of solids, by Frenkel [18] on the strength of crystalline metals, and by Mott [19] on shear-assisted slip at grain boundaries. This theory was able to predict the measured maximum strengths of a wide range of metallic systems, including pure metals [9], dilute alloys like Ni-W [9] and SmCo$_5$ [20], metallic glasses [10], and high-entropy (multi-principal-element) alloys.

Expanding on this earlier work, a generalized form is presented below that can be applied to any metallic system, still without fitting parameters. The model uses a parameterization of the initial structural and chemical disorder (**Fig. 1**), and either average-bond strength (accurately estimated for chemically-random solutions and metallic glasses) or site-specific bond strengths for chemically-ordered (intermetallic) compounds. In short, the free-energy difference between the initially partial or fully-ordered and final, fully-disordered (supercooled liquid-like) states is used to calculate the energy per unit volume or stress required to overcome it, and this determines the maximum attainable strength of an alloy.

The activation energy for this transition is the energy required to initiate the formation of a fully-disordered (amorphous) interphase with liquid-like bond density (i.e., the first or nearest-neighbors coordination number, $Z$), approximately $Z_{liq}$ = 10.5 ± 1.2 for liquid metals [21,22]. Below we show how to calculate this free-energy change for differing initial values of structural and chemical ordering via a quasi-chemical model, using changes in average bond energy and bond density. The results combine with a stress-assisted empirical rate equation for thermally-activated process (**Eq. 1**) to give a strength prediction (**Eq. 2**).

$$\dot{\varepsilon}(T) = \dot{\varepsilon}_0 \ exp\left[-\frac{\Delta G(T) - \tau V}{RT}\right]$$
Eq. 1

Here, $\dot{\varepsilon}(T)$ is the applied strain rate, $\dot{\varepsilon}_0$ is a pre-exponential factor, $\Delta G(T)$ is the molar activation free energy (J/mol), $\tau$ is the applied deviatoric shear stress, $V$ is the activation volume, $R$ is the gas constant, and $T$ is the temperature. As explained elsewhere [9], the maximum or intrinsic strength of an alloy corresponds to where the ratio of strain rate and pre-exponential factor approach a value of one. While **Eq. 1** allows a determination of strain-rate-dependent strength, implying a limit to the deformation rate that can be accommodated, we focus on the simplified case with an analytical solution corresponding to the upper strength limit. **Eq. 1** is then simplified, where an activation free energy per unit volume is adopted, $\Delta \bar{G}(T)$ (J/m³), by defining the activation volume, $V$, as the atomic volume (estimated as the ratio of liquid density and molar mass), resulting in a temperature-dependent Gibbs free energy per unit volume.

$$\tau_{max}(T) = \frac{\Delta G(T)}{V} = \Delta \bar{G}(T)$$
(Eq. 2)

An energy change diagram of the proposed concept is provided in **Fig. 2A** for the idealized case of an initially defect-free, close-packed crystalline solid. As pictured, the initial condition is a perfect crystal, but more complex structures can be treated by including parameters that describe initial structural and chemical disorder. A disordered initial state results in a reduced activation energy, and thus reduced maximum strength.

In what follows, we present reference cases for high/low structural and chemical disorder associated with the principal classes of alloys. First, a description is presented for the determination of the activation energy for pure metals, then for alloys with varying initial structural and chemical disorder, including high-entropy alloys (structurally-ordered and chemically-disordered), metallic glasses (structurally- and chemically-short-ranged-ordered), and intermetallic compounds (structurally- and chemically-ordered).

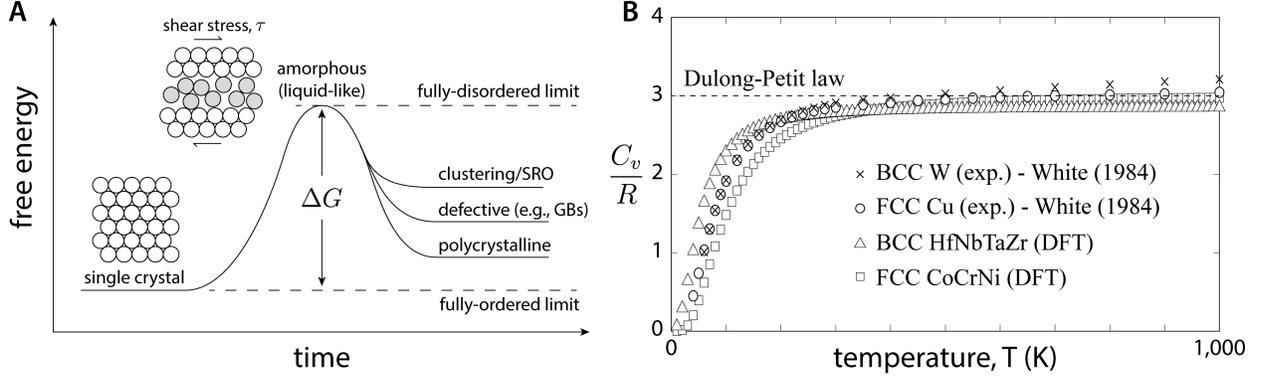

*Figure 2. (A) Schematic representation of the activation energy for flow via the formation of a fully-disordered interface, and (B) example of molar specific heat capacity for pure metals and concentrated solutions, including experimental data for pure metals from literature* [23] *and first-principles calculations for FCC (CoCrNi) and BCC (HfNbTaZr) high-entropy alloys.*

## II. Pure Metals

A thermodynamically reversible pathway (**Eq. 3**) can be used to describe the molar free-energy change, $\Delta G_{rev}(T)$, needed to transition a metal atom from a fully-coordinated (crystalline) to a supercooled liquid-like structure. The energy is related to measured and calculated values for high-energy grain boundaries, i.e., those energetically close to a liquid-like structure [9,24]. The enthalpic and entropic changes in the system are shown for each step (*i*) in a reversible energy pathway, where $C_{p,o}$ and $C_{p,d}$ are the structurally ordered (solid) and disordered (flowing) molar specific heat capacities, respectively, *L* is the molar enthalpy of fusion, and $V_{atom}$ is the atomic volume at the melting temperature, $T=T_m$.

$$\Delta G_{tot}(T) = \sum_{i=1,2,3}^{N} \Delta H_i - T \sum_{i=1,2,3}^{N} \Delta S_i$$
Eq. 3

**Step 1**: Changing temperature of the initially ordered interface from T to $T_m$ (melting),

$$\Delta H_1(T) = \int_T^{T_m} C_{p,o}(T') dT' \quad, \quad \Delta S_1(T) = \int_T^{T_m} \frac{C_{p,o}(T')}{T'} dT'$$
Eq. 4

**Step 2**: Melting the interface is a first-order transition,

$$\Delta H_2(T_m) = L \quad, \quad \Delta S_2(T_m) = L \cdot T_m^{-1}$$
Eq. 5

**Step 3**: Changing temperature of the disordered interface from $T_m$ to T,

$$\Delta H_3(T) = \int_{T_m}^{T} C_{p,d}(T')\,dT' \qquad , \qquad \Delta S_3(T) = \int_{T}^{T_m} \frac{C_{p,d}(T')}{T'}\,dT' \qquad \text{Eq. 6}$$

For pure metals, the total change in free energy is simplified using the following assumptions:

1. The molar specific heat for crystalline metals (see **Fig. 2B**) does not significantly change as a function of defect density (structural disorder) or coordination number while in the solid state [25].
2. The flowing undercooled liquid-like structure has a molar specific heat nearly identical to crystalline values, as shown by Sommer [26] and Busch *et al*. [25].

The limiting or maximum strength of pure metals corresponds to the bounding case where the shear-rate ratio in **Eq. 1** approaches a value of 1. The total free-energy change per unit volume, $\Delta \bar{G}_{tot}(T)$, with units equivalent to stress, is for an ordered-to-disordered phase transition at temperatures up to the melting temperature, as shown Eq. 7 [9].

$$\tau_{max}(T) = \Delta \bar{G}_{tot}(T) = \frac{L}{V_{atom}}\left(1 - \frac{T}{T_m}\right) \qquad \text{Eq. 7}$$

This was shown to accurately describe strength limits in pure metals [9]. **Figure 3** shows a comparison of predicted maximum strengths and those reported from MD simulations and experiments for a variety of pure elements. Experiments include uniaxial tensile-flow strengths and indentation hardness, all converted to shear strengths, using the relationship between shear ($\tau$), uniaxial tension ($\sigma$), and hardness (*H*), $H \approx 3\,\sigma \approx 3\sqrt{3}\,\tau$, via Tabor and von Mises criteria, respectively. High strength is achieved by inhibiting dislocation activity through the presence of regions of structural disorder, e.g., grain boundaries (GBs). In most cases the GB are regions of high disorder, sometimes structurally near-amorphous [9,24], and this reduces the activation energy for flow. For randomly textured polycrystalline systems with grain sizes smaller than approximately *d* = 10-30 nm, the initial state of structural disorder significantly reduces the activation energy for flow. A simple approximation can be used to account for this energy reduction [9] by using the volume fraction of crystalline material (as shown in Eq. 8) and assuming that the GBs are effectively close enough to amorphous to be of negligible import.

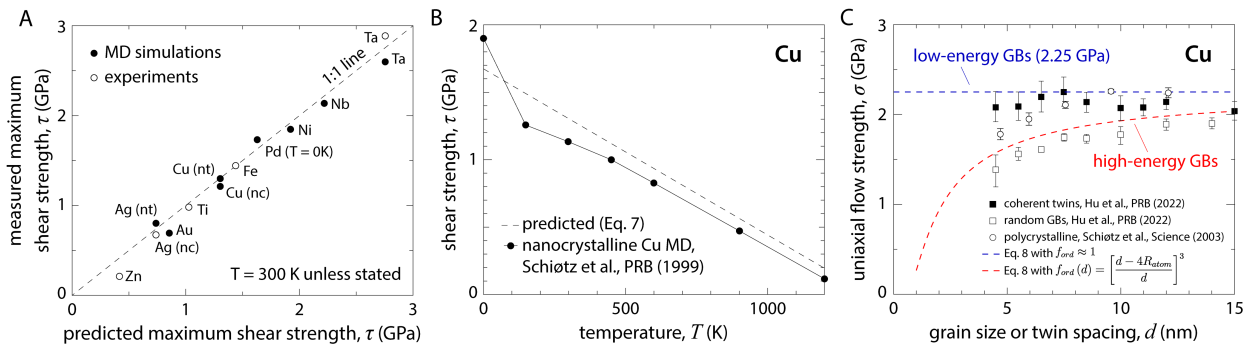

*Figure 3. Comparison of (A) predicted maximum strengths for pure metals compared to data from experiments and MD for pure metals Zn [27], Ag [28,29], Au [6], Ti [30], Cu (nc / nt refer to*

*nanocrystalline / nanotwinned microstructures)* [5,31], *Fe* [30], *Pd* [32], *Ni* [33], *Nb* [33], *and Ta* [34,35], *(B) temperature-dependent change in strength for Cu from MD by Schiøtz et al.* [32], *and (C) twin spacing- and grain size-dependent uniaxial flow stress determined for Cu by Hu et al.* [31] *and Schiøtz et al.* [5].

Overlays of predicted behavior and published data shown in **Fig. 3A-B** assumes that these systems are initially perfectly ordered, with an initially structurally-ordered volume fraction of one, $f_{ord}$ = 1, a limiting case in Eqs. 8. For example, the effect of GBs, which are a dominant factor in the nanocrystalline regime, are neglected in these predictions. For grain sizes $d > 10$ nm, the volume fraction of grain interior (structurally-ordered) to grain boundary (structurally-disordered) is $f_{ord} > 0.85$, so this assumption introduces a small error compared to factors like experimental uncertainty in average grain size determination. An example of where this assumption becomes significant and thus problematic is shown in **Fig. 3C**, with data from Hu *et al.* These authors showed that it is possible to retain the predicted maximum strength in Cu at grain sizes below which softening is typically observed with randomly textured polycrystalline metals. This was accomplished by exclusively introducing coherent twins, essentially low free-energy GBs, rather than randomly oriented and highly-disordered GBs like those in most earlier simulations and experiments. It is notable that their demonstration of a strength plateau in Cu for coherent twin spacing below 10 nm closely matches the predicted maximum strength. This result serves as validation of the proposed method, and the approach of including the effects of initial structural disorder through the volume fraction ($f_{ord}$) of structurally-disordered material [36,37], e.g., the volume fraction of high-angle GBs. In this case, where $d$ is the average grain size and $\delta$ is the GB width, approximately twice the average atomic diameter or width of two atomic planes [24],

$$\tau(d,T) = \bar{G}(T) f_{ord}(d) \qquad \text{Eqs. 8}$$

where,

$$f_{ord} = \left(\frac{d-\delta}{d}\right)^3$$

### III. Structurally-Ordered/Chemically-Disordered Alloys (e.g., High-Entropy Alloys)

The same analysis can be applied to alloys by using a regular-solution model, combining rule-of-mixtures (ROM) based on the elemental constituents and adding a correction for mixing free energy. The free energy change requires a contribution from the free energy of mixing, which can be introduced in a similar way to how Miracle et al. [38] showed it is possible to estimate or predict heats of formation, enthalpies of fusion, and surface energies for intermetallic compounds, as well as the following additional assumptions:

1. The excess molar specific heat at constant pressure ($C_P$) for disordered alloys (e.g., HEAs and BMGs) is proportional to the free energy of mixing ($\Delta G_{mix}$); this assumption was validated by Witusiewicz and Sommer for binary mixtures [39]. Sommer also provide an analytical model for estimating the excess specific heat of miscible and immiscible alloys [26].

2. For fully chemically-disordered alloys, including dilute and concentrated alloys (MPEAs and HEAs), measurements and calculations of molar specific heats were shown to follow the Dulong-Petit law, converging at high temperatures to $C_P$ = 2.9-3.4R. Examples include single-phase $Al_{0.3}CoCrFeNi$ and CoCrFeNi, both shown to have a measured molar specific heat of $C_P$ = 2.98R at room temperature [40,41], and equimolar refractory HEAs HfNbZr, HfNbTiZr, and HfNbTaTiZr, with values in the range $C_P$ = 2.9-3.4R [42].

The activation energy per unit volume for the generation of an amorphous or structurally fully-disordered interface in a structurally-ordered, chemically-disordered alloy is a sum of the ideal average ($\Delta \bar{G}_{ideal}$ from rule-of-mixtures) and a correction for mixing free energy ($\Delta \bar{G}_{reg}$).

$$\Delta \bar{G}_{tot}(T) = \Delta \bar{G}_{ideal}(T) + \Delta \bar{G}_{reg}(T) \qquad \text{Eqs. 9}$$

where,

$$\Delta \bar{G}_{ideal}(T) = \sum_{i=1,2...}^{n} x_i \Delta \bar{G}_i(T)$$

$$\Delta \bar{G}_{reg}(T) = \left(1 - \frac{z_{mix}}{z_{liq}}\right) \frac{\Delta G_{mix}(T)}{V_{atom}}$$

Here, $x_i$ is the atomic fraction of each element, $\Delta G_{mix}$ corresponds to the mixing free energy, $z_{mix}$ is the average coordination number in the initial solid phase, and $V_{atom}$ is the average atomic volume of the alloy. For the examples below, $\Delta G_{mix}$ was calculated with density-functional theory (DFT). Few HEA datasets exist with which to validate predictions at the Hall-Petch breakdown limit, and those that are available are exclusively from atomistic simulations using interatomic potentials that were not developed to accurately describe properties like specific heat capacity and melting temperature. However, in **Fig. 4A** we show a comparison of published MD data [43,44] for a few systems and compare these to predicted strengths, based on **Eq. 8**, and find remarkably good agreement for the maximum strength of both FCC and BCC HEAs.

A more comprehensive but indirect method for validating the model is to compare predicted strengths with DFT-based bulk moduli. The bulk moduli were shown to be directly proportional to ultimate strength by Johnson et al. [14], with activation free energies calculated using DFT-based formation enthalpies. In **Fig. 4A** we also present a comparison of these parameters for thousands of BCC and FCC HEAs (see supplemental data).

### A. structurally-ordered/chemically-disordered metals | high-entropy alloys

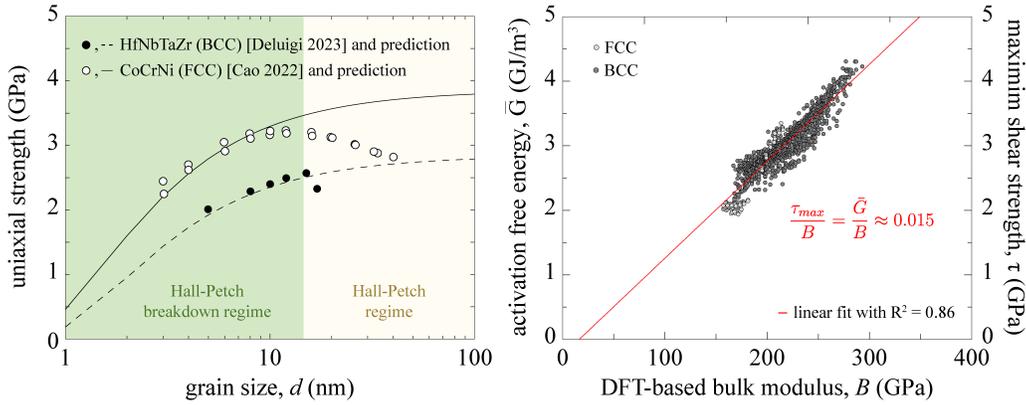

### B. structurally- and chemically-ordered metals | intermetallic compounds

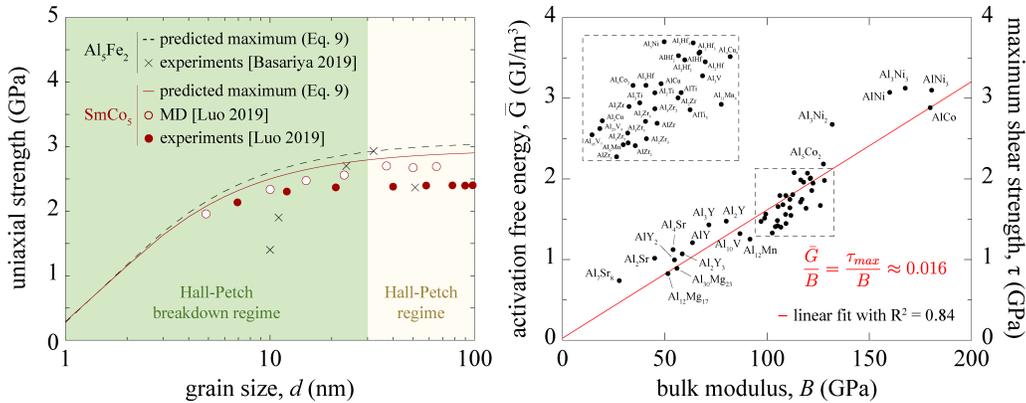

### C. structurally- and chemically-SRO metals | metallic glasses

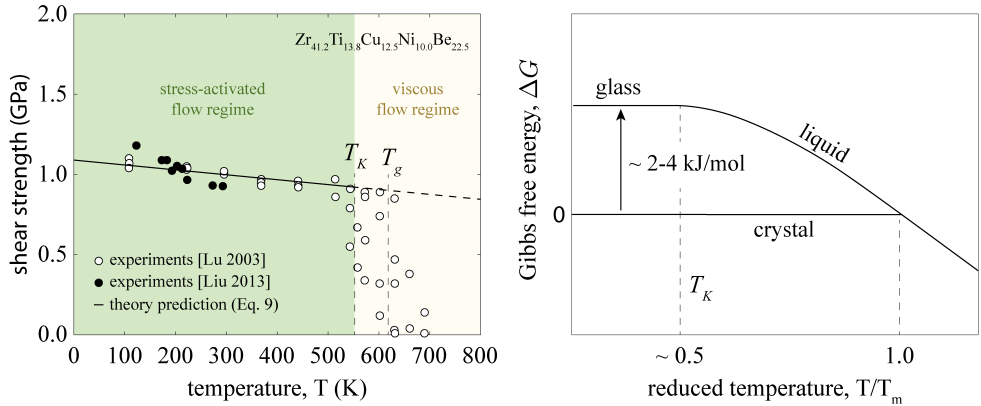

*Figure 4.* Comparison of predicted, experimental, and calculated strengths as a function of grain size and correlation of predicted strength with bulk modulus for (A) MPEAs and HEAs and (B) binary intermetallic compounds. While the theory fails to predict the grain-size dependent strength of $Al_5Fe_2$ here for reasons explained in the accompanying text, the ultimate strength, which represents an upper bound, is in good agreement. Formation energies and measured elastic moduli for the Al-based intermetallic compounds are from DFT calculations in Wang et al. [45]. (C) A representative comparison of predicted and measured shear strengths for metallic glass

*alloys, i.e., $Zr_{41.2}Ti_{13.8}Cu_{12.5}Ni_{10.0}Be_{22.5}$, adapted from Argibay and Chandross [10], and a schematic of measured temperature-dependent free-energy differences between undercooled liquid and crystalline phases for metallic glasses (adapted from Busch et al. [25,46]) showing that short-range order in glasses results in a temperature-independent free-energy offset for $T < T_K$, supporting the proposed framework.*

The comparison between DFT-based bulk moduli and maximum strengths for many alloys shows significant scatter, though with a multi-modal distribution, with clustering about a linear proportional trendline with the same slope for both FCC and BCC. One possible source of error is that the calculation of maximum strengths is based exclusively on materials properties, including the latent heats of fusion and atomic diameters for elements that may have a lattice structure in their pure form that differs from that in the alloy structure. It is likely that a correction factor can be used based on bond density or coordination number to account for differences in pure and alloy lattice structure. This is conceptually similar to what was demonstrated by Miracle et al. [38,47] for intermetallic compounds, who showed that it is possible to accurately predict formation enthalpies and other thermophysical properties using average bond energies and coordination numbers taking into account lattice site occupancy using statistical methods. Johnson *et al.* also showed that maximum strength (or moduli) can be predicted for metals, including HEAs/MPEAs, from the DFT-based equilibrium interstitial electron charge densities [14]. In that work, the lattice-specific elemental charge density values were provided as supplemental information and can be used (as was shown) for low-cost estimates and more accurate predictions of strength and ductility, ensuring that elemental volumes are set to match the lattice structure for the alloy of interest.

### IV. Structurally- and Chemically Ordered Alloys (Intermetallic Compounds)

Chemically- and structurally ordered alloys, *i.e.,* intermetallic compounds, exhibit long-range order and thus a higher probability of dissimilar elemental bond pairs resulting in larger deviations from ideality compared to chemically random alloys like high-entropy alloys. Additionally, deviations from idealized unit cell elemental distributions are common experimentally, requiring statistical descriptions of site occupancy by each elemental constituent. Deviations from equilibrium conditions can result from multiple factors including compositional nonuniformity during solidification. These deviations are difficult to measure and can result in significant errors when comparing to predictions. Published data for intermetallic compounds that probe the relevant grain-size regime, where the maximum strength limit is reached, are scarce, and ductility is limited in these materials. This implies the possibility of materials failing due to fracture at fine grain sizes, where defects can act as nucleation sites. This can occur at macroscopic stresses that are well below the maximum predicted value.

Published experimental grain-size dependent strengths in the ultra-nanocrystalline regime (grain sizes $d \cong 5\text{-}50$ nm) are compared with predictions based on Eq. 8 and 9 in **Fig. 4B**. In these examples, the mixing free energy was taken to be a DFT-based or experimentally-measured free energy of formation. **Figure 4B** shows a comparison for two alloys, one where a systematic investigation of grain-size dependent strength was carried out experimentally on $Al_5Fe_2$ [48] and both experimentally and via molecular dynamics on $SmCo_5$ [49]. In **Fig. 4B** we also present a comparison of predicted strengths and calculations for Al-based binary intermetallics using published formation free energies and average coordination numbers from Wang et al. [45] and Miracle et al. [38], respectively.

## V. Short-Range Ordered/Cluster-Forming Alloys (Metallic Glasses)

Analyses of molar specific heat capacity as a function of temperature have been published for multiple bulk metallic-glasses, including those by Busch *et al.* [25,50]. **Figure 4C** illustrates results from experimental investigations of temperature-dependent specific heat and free energy changes in metallic glasses [46]. These investigations showed that below the glass-transition ($T_g$) or Kauzmann ($T_K$) temperatures [25] the molar specific heats of metallic glasses are effectively identical to those for the crystalline phase and that alloying and SRO introduce a constant (excess) Gibbs free energy.

The excess free energy associated with formation of a fully-disordered interface can be calculated analytically for temperatures below $T_K$ [10]. Values used for the initial short-range structural ordering ($f_{ord}$) are a likely source of significant error in the calculations presented in the earlier work, which relied on work by Egami [51] and Miracle [52,53] to estimate the average bond density or coordination number change associated with the formation of a fully-disordered interface in a glassy structure. The SRO length scale was proposed to follow the form of Eq. 10 [10], where the ordered domain size (*d*) represents the size of a close-packed cell of atom clusters, ranging from 1.6-2.2 nm or average value of $d \cong 2$ nm. The close-packing of clusters was used to explain the experimental densities of metallic glasses that are similar to those for FCC crystals [52,54]. The thickness of an amorphous interface, $\langle \delta \rangle$, was defined as twice an average atomic diameter as determined from the measured density of the glassy alloy or estimated as a rule-of-mixtures of the atomic diameters of the coordinated (crystalline) constituent elements [55].

$$f_{ord} = \left( \frac{d_{cell} - \langle \delta \rangle}{d_{cell}} \right)^3$$

Eqs. 10

where,

$$\langle \delta \rangle = \sum_{i=1,2...}^{n} x_i d_i$$

These approximations resulted in reasonably good agreement with measurements of temperature-dependent strength for a large number of metallic glasses [10]. An experimental determination of coordination number, via high-energy Debye X-ray scattering, provides a direct reference to validate assumptions about SRO dimensions and predictions. A pair distribution analysis using

Voronoi tessellation was used to determine a coordination number distribution for $Zr_{41.2}Ti_{13.8}Cu_{12.5}Ni_{10}Be_{22.5}$ [56], which gave an average value of $z_{ord}$ = 15. The free energy of mixing based on the method provided by Takeuchi and Inoue [57] was used to calculate the temperature-dependent shear strength, and a comparison between experimental data [58,59] is presented in **Fig. 4C**.

**Summary and Conclusions**

A quantitative energy-based relationship between order and strength was established for pure metals and alloys, requiring no fitting parameters, and based exclusively on materials properties and structural and chemical ordering parameters. This relationship was shown to be universally valid for a wide range of structurally and chemically disordered metals and linked to charge density-based predictions that accurately predict maximum strength, stiffness, and ductility in high-entropy alloys. Notably, this work likely can be expanded to establish direct correlations with the activation energy for dislocation emission and glide, and, by extension, also fracture and ductility. This work demonstrates the existence of a predictable limiting case of strength for all kinds of metals based on the activation energy for localized stress-assisted amorphization. Linking this work with CALPHAD and rapid [60] DFT methods can enable a robust framework for alloy design that includes balancing tradeoffs between strength and ductility.

**Acknowledgments**

This work was funded by the Laboratory-Directed Research and Development (LDRD) program at Ames National Laboratory. This article has been co-authored by an employee of National Technology & Engineering Solutions of Sandia, LLC, under Contract No. DE-NA0003525 with the U.S. Department of Energy (DOE). The employee owns all rights, title, and interest in and to the article and is solely responsible for its contents. The publisher, by accepting the article for publication, acknowledges that the United States Government retains a non-exclusive, paid-up, irrevocable, world-wide license to publish or reproduce the published form of this article or allow others to do so, for United States Government purposes. The DOE will provide public access to these results of federally sponsored research in accordance with the DOE Public Access Plan https://www.energy.gov/downloads/doe-public-access-plan.

**Methods and Supplemental Material**

*DFT Molar Specific Heat Capacity* | The molar specific heat capacities of CoCrNi (108 atoms; 3x3x3 of conventional FCC unit-cell) and HfNbTaZr (72 atoms; 3x3x4 of conventional BCC unit-cell) were calculated using density-functional perturbation theory (DFPT) method, as implemented in the *Vienna Ab-initio Simulation Package* (VASP) [61–63] combined with phonopy [64]. The atomic displacement of 0.04 Å was used for all resutls. We employed plane-wave basis projector augmented wave method [65] and Perdew-Bueke-Ernzerhof (PBE) [66] exchange-correlation functional in the generalized gradient approximation [65]. A plane-wave

energy cutoff of 520 eV and sufficiently dense k-point mesh (5x5x5/CoCrNi; 5x5x3/HfNbTaZr) to realize full relaxation of energy ($10^{-6}$ eV/cell) and force ($10^{-6}$ eV/A).

*DFT Mixing Energy* | A DFT-based Green's function, all-electron Korringa-Kohn-Rostoker (KKR) method [67] in combination with the coherent-potential approximation (CPA) was employed to provide the configurationally averaged results for homogeneous random alloys, including the effect of Friedel-like screening in the random alloy. The generalized gradient-corrected (PBEsol) exchange-correlation functional [68] was used throughout to estimate mixing free energy (formation enthalpy), as implemented through *LibXC* package.